\RequirePackage{fix-cm}
\documentclass[twocolumn,epjc3]{svjour3}  
\smartqed  
\RequirePackage{graphicx}
\usepackage{xcolor}
\usepackage{amsmath}
\usepackage{amssymb}
\usepackage{soul}
\usepackage{xspace}

\newcommand{\eg}{\textit{e.g.}\xspace}
\newcommand{\ie}{\textit{i.e.}\xspace}

\newcommand{\reffig}[1]{\textrm{Fig}.~\ref{#1}}

\journalname{Eur. Phys. J. A}
\begin{document}

\title{Impact of a positron beam at JLab on an unbiased determination of DVCS Compton Form Factors}

\author{
H.~Dutrieux \thanksref{email1,address1} \and
V.~Bertone \thanksref{email2,address1} \and
H.~Moutarde \thanksref{email3,address1} \and
P.~Sznajder \thanksref{email4,address2}
}

\thankstext{email1}{e-mail: herve.dutrieux@cea.fr}
\thankstext{email2}{e-mail: valerio.bertone@cea.fr}
\thankstext{email3}{e-mail: herve.moutarde@cea.fr}
\thankstext{email4}{e-mail: pawel.sznajder@ncbj.gov.pl}

\institute{%
IRFU, CEA, Universit\'e Paris-Saclay, F-91191 Gif-sur-Yvette, France \label{address1}
\and
National Centre for Nuclear Research (NCBJ), Pasteura 7, 02-093 Warsaw, Poland \label{address2}
}

\date{Received: date / Accepted: date}

\maketitle

\sloppy

\begin{abstract}
The impact of potential future measurements of beam charge asymmetries on the current knowledge of Compton form factors is evaluated. 
Ela\-bo\-rating on the results of a global neural network fit to deeply virtual Compton scattering data, a Bayesian reweighting analysis quantifies the improved determination of the real part of the Compton form factor $\mathcal{H}$.
Such an improvement is particularly relevant for the experimental determination of the proton me\-cha\-nical properties or for universality studies of ge\-ne\-ra\-lized parton distributions through analyses of virtual Compton scattering in both spacelike and timelike regions.
\end{abstract}


%
%
\section{Introduction}

Introducing the concept of generalized parton distributions (GPDs) \cite{Mueller:1998fv,Ji:1996ek,Ji:1996nm,Radyushkin:1996nd,Radyushkin:1997ki} immediately triggered an intense experimental interest for hard exclusive experiments, and in particular for deeply virtual Compton scattering (DVCS) \cite{Airapetian:2001yk,Stepanyan:2001sm}. 
GPDs provide detailed information about the three-dimensional partonic structure of the nucleon with precious insights into the spin structure, charge distribution, mechanical properties or even the perspective of partonic tomography.
Theoretical progress \cite{Ji:1998pc,Goeke:2001tz,Diehl:2003ny,Belitsky:2005qn,Boffi:2007yc,Guidal:2013rya,Mueller:2014hsa} has been accompanying experimental programs \cite{dHose:2016mda} over the last two decades and keeps on enriching the physics case for future facilities \cite{Accardi:2012qut,AbdulKhalek:2021gbh,Anderle:2021wcy,AbelleiraFernandez:2012cc}. 

Compton form factors (CFFs) encode information about 3D nucleon structure that can be probed through DVCS and constitute an intermediate step between GPDs and DVCS measurements. 
The determination of CFFs from experimental data is usually attempted either through local fits (independent extractions of CFFs on each kinematic bin) or through global fits (simultaneous extraction of CFFs on a large set of kinematic bins) (see Ref.~\cite{Kumericki:2016ehc} and references therein).
More sophisticated extractions of CFFs recently combined local fits with kinematic corrections \cite{Defurne:2017paw,Benali:2020vma}, Rosenbluth separation \cite{Kriesten:2020apm} or even para\-me\-tric global fits \cite{Dupre:2016mai,Dupre:2017hfs,Burkert:2018bqq,Burkert:2021ith}. 
Most of the existing DVCS measurements have been included in global fits of CFFs, either using a physically-motivated parameterization \cite{Moutarde:2018kwr} or artificial neural networks (ANNs) \cite{Kumericki:2019ddg,Cuic:2020iwt,Moutarde:2019tqa}. These global fits provide a useful reference to move forward. In particular, some fit results are publicly available as software libraries and can be used for new phenomenology stu\-dies.

So far, DVCS has mostly been measured with an electron or a muon beam scattering off a proton target. 
Only the experimental settings of HERMES and COMPASS permitted a change of the beam charge, brin\-ging unique information, yet over a restricted kinematic region where sea quarks are expected to dominate.
This situation can dramatically change with the possibility of operating a polarized positron beam~\cite{Afanasev:2019xmr,Accardi:2020swt} at the Thomas Jefferson National Accelerator Facility (JLab) to similarly probe the valence region.
We study here the impact of such potential DVCS data in the kinematic reach of the CLAS12 spectrometer~\cite{Burkert:2020akg}. 
We conduct a Bayesian reweighting analysis to evaluate how these data would improve the determination of CFFs obtained from a global fit with ANNs \cite{Moutarde:2019tqa}. 
Reweighting techniques \cite{Ball:2010gb,Ball:2011gg} are now common in the phenomeno\-logy of parton distribution functions (PDFs) and are even used to quantify the impact of lattice QCD calculations on the global determination of PDFs (see \eg Ref.~\cite{Lin:2017snn}). 
To the best of our knowledge, the simultaneous recourse to a neural network global fit of CFFs and Bayesian reweighting is a new original approach in the context of GPD studies with a positron beam. 

The following section describes the CFF determination used for the present study. Then we briefly motivate beam charge dependent observables based on the structure of the DVCS cross-section. We continue with details on the generation of pseudo-data and the implementation of the Bayesian reweighting strategy. At last, we provide our results and draw conclusions.

\section{Description of input Compton form factors}

CFFs are complex-valued functions of the virtuality $Q^2$ of the photon mediating the interaction between the lepton beam and the proton target, of the Mandelstam variable $t$ associated to the four-momentum transfer to the target, and of the fractional longitudinal momentum transfer $\xi$ to the active parton.
The present study is based on extractions of DVCS CFFs obtained in a global analysis of experimental data collected with proton targets \cite{Moutarde:2019tqa} encompassing about 30 distinct DVCS observables measured in more than 2600 kinematic points. 
The publication of these data by the Hall-A, CLAS, HERMES, COMPASS, ZEUS and H1 experiments spans over 17 years and provides a wide phase-space coverage. 
The additional conditions 
\begin{eqnarray}
   Q^{2} & > & 1.5~\mathrm{GeV}^2 \,, \label{eq:constrainQ2} \\
    -t & < & 0.2\,Q^{2} \,, \label{eq:constraint}
\end{eqnarray}
allow us to avoid significant higher-twist corrections. 
This CFF extraction is based on ANNs and should be free of the model bias affecting global analyses based on classical Ans\"atze. 
The CFF extraction from DVCS measurements is ``agnostic’’: real and imaginary parts of CFFs are described by separate ANNs, a full kinematic dependence of CFFs is taken into account, no power-law pre-factors are used. 
As a result, the obtained CFFs convey a refined information coming from global experimental data, making them an ideal input to impact studies.  

The propagation of experimental uncertainties is done with the so-called replica method, which is at the heart of the reweighting procedure. 
In the replica method, the fit is repeated $N_{\mathrm{rep}}$ times, each time randomly smearing experimental points around their mean values according to the associated uncertainties. 
As a result of such a procedure, one obtains $N_{\mathrm{rep}}$ parameterizations of CFFs, referred to as replicas (here $N_{\textrm{rep}} = 101$).
In this original set, at a specific $(\xi, t, Q^{2})$ kinematic configuration, some replicas are identified as outliers and discarded using the $3\sigma$-method \cite{Dutrieux:2021nlz}. This yields a reduced set of $N'_{\mathrm{rep}} \leq N_{\mathrm{rep}}$ replicas at each kinematic configuration.
These replicas can then be used for instance to estimate the mean, $\mu$, and standard deviation, $\sigma$, at the considered kinematic point (here the example for $\mathrm{Im}\,\mathcal{H}$):
\begin{equation}
\label{eq:def-replica-mean}
  \mu_{\mathrm{Im}\,\mathcal{H}}(\xi, t, Q^{2}) = 
	\frac{1}{N'_{\mathrm{rep}}}
	\sum_{k = 1}^{N'_{\mathrm{rep}}} \mathrm{Im}\,\mathcal{H}_{k}(\xi, t, Q^{2}) \,, 
\end{equation}
\begin{flalign}
\label{eq:def-replica-variance}
  \sigma_{\mathrm{Im}\,\mathcal{H}}^{2}&(\xi, t, Q^{2}) = \nonumber \\
	&\frac{1}{N'_{\mathrm{rep}}}
	\sum_{k = 1}^{N'_{\mathrm{rep}}} \left( \mathrm{Im}\,\mathcal{H}_{k}(\xi, t, Q^{2}) - \mu_{\mathrm{Im}\,\mathcal{H}}(\xi, t, Q^{2})  \right)^2\,,
\end{flalign}
where $\mathrm{Im}\,\mathcal{H}_{k}(\xi, t, Q^{2})$ is the value associated to the $k$-th replica.

\section{Beam charge dependent observables at JLab}

We adopt the notations of Ref.~\cite{Kroll:2012sm} for DVCS observables. The differential cross-section for the single photon lepto-production off an unpolarized proton target, $l p \rightarrow l' p' \gamma$, is:
\begin{align}
\label{eq:DVCSGeneralCrossSection}
\frac{\mathrm{d}^4\sigma^{h_l, e_l}}{
\mathrm{d}x_{\mathrm{B}} \,
\mathrm{d}t \,
\mathrm{d}Q^{2} \, 
\mathrm{d}\phi
} = \frac{\mathrm{d}^4\sigma_{\textrm{UU}}}{
\mathrm{d}x_{\mathrm{B}} \,
\mathrm{d}t \,
\mathrm{d}Q^{2} \, 
\mathrm{d}\phi
} \big[ 1 + e_l A_{\textrm{C}}(\phi) \\ \nonumber 
\quad  + h_l A_{\textrm{LU, DVCS}}(\phi) +  e_l h_l A_{\textrm{LU, I}}(\phi)  \big]\,,
\end{align}
where $h_l/2$ denotes the helicity of the beam particle. 
$e_l$ is the beam charge in units of the positron charge $|e|$. 
$\phi$ is the angle between the leptonic plane (spanned by the incoming and outgoing lepton momenta) and the production plane (spanned by the virtual and outgoing photon momenta).
$x_{\mathrm{B}}$ relates to the longitudinal momentum transfer $\xi$ through $\xi = x_{\mathrm{B}} / (2-x_{\mathrm{B}})$.
This cross-section gets contributions from DVCS, but also from Bethe-Heitler (BH), which is the electromagnetic process with the same initial and final states as DVCS, and the interference between DVCS and BH.  
For concision, only the $\phi$-dependence of the observables is shown.
The cross-section expressed in Eq.~\eqref{eq:DVCSGeneralCrossSection} can be parametrized in terms of elastic form factors (EFFs) and CFFs, as described for instance in Ref.~\cite{Belitsky:2012ch}. 
At a fixed kinematic point, different combinations of CFFs are probed through different experimental settings like the polarisations and charges of colliding particles. 
Measuring cross-sections varying these settings thus provides a leverage to probe a specific CFF, which can be the key to access valuable information about GPDs. 
This motivates the specific structure of Eq.~\eqref{eq:DVCSGeneralCrossSection}. The beam charge asymmetry (BCA) $A_{\textrm{C}}$ can be obtained from the combination: 
\begin{equation}
A_{\textrm{C}}(x_{\mathrm{B}}, t, Q^{2}, \phi) = \frac{
\mathrm{d}^4\sigma^{+} - \mathrm{d}^4\sigma^{-} 
}{
\mathrm{d}^4\sigma^{+} + \mathrm{d}^4\sigma^{-} 
} \,,
\end{equation}
where $\mathrm{d}^4\sigma^{\pm}$ denotes the single photon lepto-production cross-sections averaged over the helicity of the beam particle (\ie measured for unpolarized beams and targets), for either positively $(+)$ or negatively $(-)$ charged beam particles.

The BCA $A_{\textrm{C}}$ is primarily sensitive to the real part of CFFs, in particular to that of the CFF $\mathcal{H}$ (see the orders of magnitude of Ref.~\cite{Kroll:2012sm}). 
As an illustration, we remind the following (leading twist and leading order) relation between CFFs and the $\cos \phi$ Fourier harmonics of $A_{\textrm{C}}$:
\begin{equation}
A_{C}^{\cos{\phi}} \propto
\mathrm{Re}\left[
F_{1}\mathcal{H} + 
\xi\left( F_{1} + F_{2} \right)\widetilde{\mathcal{H}} - 
\frac{t}{4M^{2}}F_{2}\mathcal{E}
\right] \,,
\end{equation}
where $F_{1}$ and $F_{2}$ are respectively the Dirac and Pauli EFFs, and where $M$ stands for the proton mass. 
Because of this relation, measurements of BCA are widely considered to be one of the best sources of information about the real part of the CFF $\mathcal{H}$. 

A precise BCA measurement requires an accelerator facility operating with lepton beams of both charges ($e^{\pm}$ or $\mu^{\pm}$) and high luminosities, and an experimental apparatus capable of reconstructing the entire final states of events, including electromagnetic contributions. These conditions may be met by JLab after equipping the Continuous Electron Beam Accelerator Facility (CEBAF) with a positron source of high  efficiency. In this work, we investigate the foreseen impact of measurements of the BCA $A_{\textrm{C}}$ at CLAS12 on the current knowledge of CFFs.

\section{Impact analysis settings}

\subsection{Generation of pseudo data}

\begin{figure}
\includegraphics[scale=0.95]{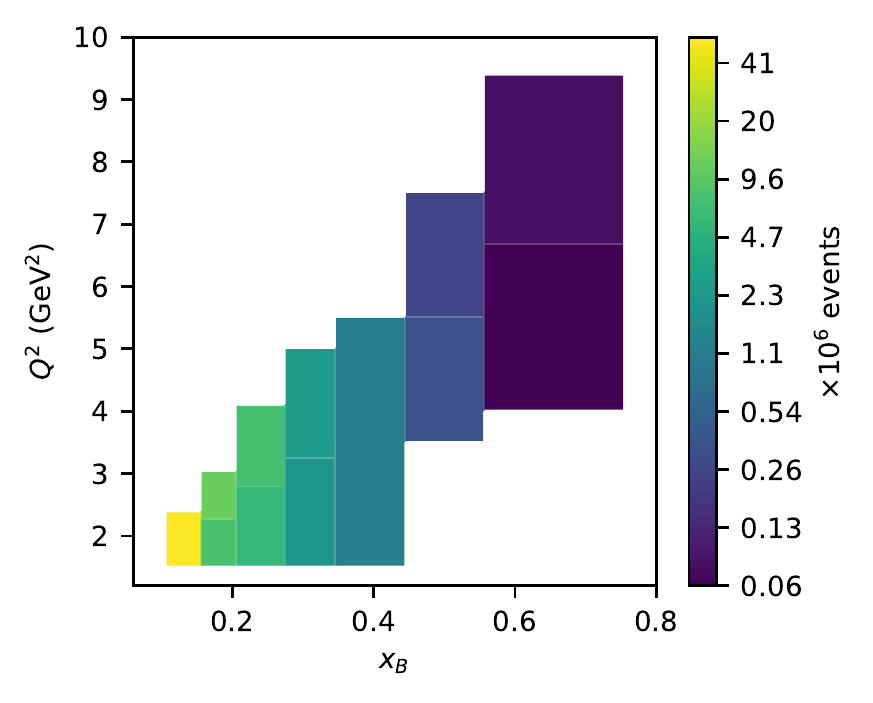}
\caption{\label{fig:nevents} Number of DVCS/BH events off unpolarized protons for unpolarized positron and electron beams sharing 80 days of data taking, integrated over the $t$ and $\phi$ bins for the thirteen $(x_B, Q^2)$ bins. The beam has an energy $E = 10.6$~GeV and a luminosity $\mathcal{L} = 0.6\times 10^{35}$~cm$^{-2}\cdot $s$^{-1}$. A perfect detector acceptance and efficiency is assumed. The result is obtained using the physical fitting Ansatz of Ref.~\cite{Moutarde:2018kwr}.}
\end{figure}

We assume that BCAs will be measured in the CLAS12 spectrometer with CEBAF operating in both electron and positron modes at $10.6~\mathrm{GeV}$. 
We select 13 bins in $(x_B, Q^2)$ as shown in \reffig{fig:nevents}. 
Each of these bins is further divided into at most 6 bins in $t$ and 24 bins in $\phi$ spanning the range $[0, 2\pi]$. Enforcing conditions \eqref{eq:constrainQ2} and \eqref{eq:constraint}, 1656 kinematic bins cover the $(x_B, Q^2, t, \phi)$ phase space. 
For each of these kinematic configurations, we simulate the BCA $A_C$ produced by the expected JLab positron and electron beams sharing 80 days of data taking. 
We denote with $\langle A_{C,i} \rangle$ the mean of $A_C$ evaluated over the ANN replicas of the CFFs at the central kinematics of a given bin $i$ computed using Eq.~\eqref{eq:def-replica-mean}.
We simulate the uncertainty of expected future experimental points as
\begin{equation}
    \Delta A_{C,i} = \sqrt{\frac{1-\langle A_{C,i}\rangle^2}{N_i}} \oplus 0.03\, \langle A_{C,i}\rangle\,,
\end{equation}
where $\oplus$ denotes quadrature addition ($a \oplus b = \sqrt{a^2 + b^2})$, an uncorrelated 3\% systematic uncertainty has been added \cite{burkert:pac48}, and $N_i$ is the number of expected BH/DVCS events in the bin.  
This number of events is given by the product of the luminosity, data taking duration, differential cross-section and phase space volume element. 
We assume perfect detector acceptance and efficiency, and an expected luminosity of $\mathcal{L} = 0.6\times10^{35}\, \textrm{cm}^{-2}\cdot\textrm{s}^{-1}$.
Finally, with both $\langle A_{C,i}\rangle$ and $\Delta A_{C,i}$ evaluated, we take as the central value of the expected future experimental points the mean value of $A_C$ evaluated over the ANN replicas smeared by the expected uncertainty, 
\begin{equation}
    \label{eq:drawcentral}
    \mathrm{rnd}\big(\langle A_{C,i}\rangle, \Delta A_{C,i}\big) \pm \Delta A_{C,i} \,,
\end{equation}
where $\mathrm{rnd}(\mu, \sigma)$ represents a random variable following the normal distribution characterised by its mean $\mu$ and standard deviation $\sigma$. Such smeared values, mimicking the outcome of a real measurement, are used in the next part of the analysis.

To estimate the differential BH/DVCS cross-section, we do not rely on the ANN extraction of CFFs, but instead on a physically-motivated parameterization constrained in a global CFF fit \cite{Moutarde:2018kwr}. 
The number $N_i$ of events expected in each $(x_B, Q^2)$ bin, integrated over the $t$ and $\phi$ bins, is shown in \reffig{fig:nevents}.
It is a feature of ANN fits that CFFs are weakly constrained in regions where little to no data are available. 
In particular, the used ANN extractions of CFFs are essentially consistent with zero in the large $(x_B, Q^2)$ region. 
Since the $l p \rightarrow l' p' \gamma$ cross-section is a non-linear positive function, the ANN estimates of the number of events present a very distorted distribution for these kinematics (peaked at small values and with a heavy tail). 
Such a distribution tends to considerably overestimate the actual number of events, especially at large $t$. 
\textit{A contrario}, a phenomenological model or a physical fitting Ansatz will typically see a quick reduction of the differential cross-section at large $t$.

\begin{figure}
\includegraphics[scale=0.75]{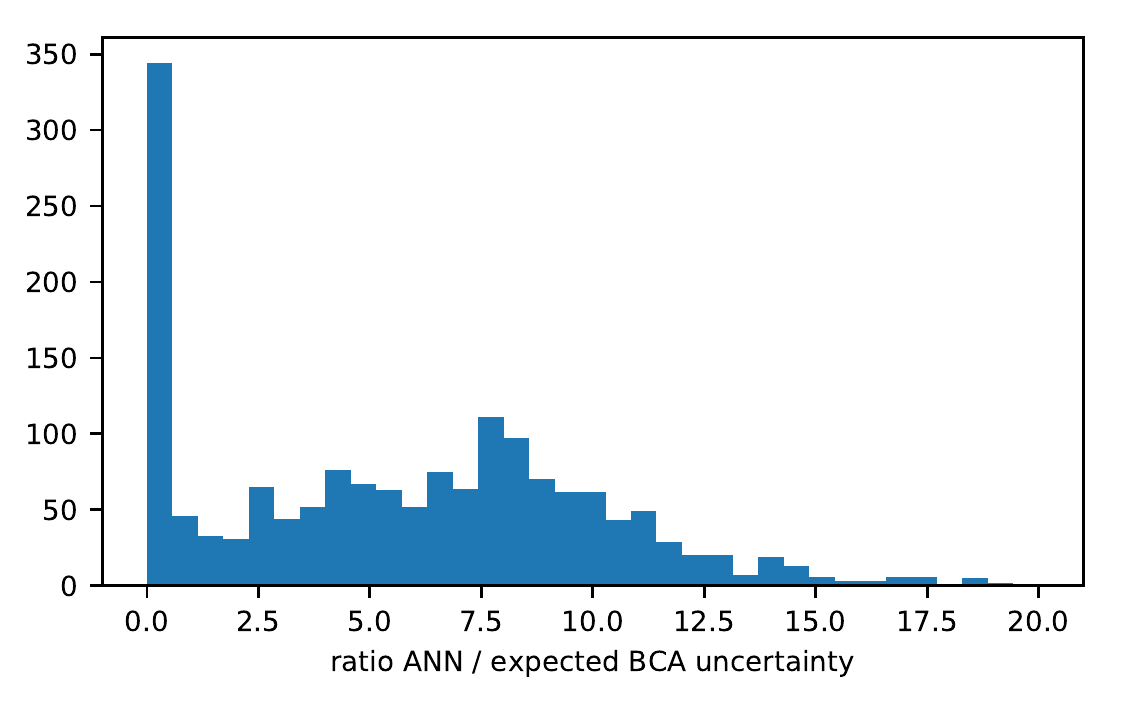}
\caption{\label{fig:ratiostd} Distribution of the ratio of the current uncertainty on the unpolarized BCA with ANN CFFs divided by the expected experimental uncertainty on the 1656 cinematic bins used in this study.}
\end{figure}

\reffig{fig:ratiostd} shows the distribution over the 1656 kinematic configurations of the ratio of the BCA uncertainty estimated with ANN CFFs over the BCA uncertainty expected from JLab measurements with a positron beam. 
Many kinematics come with an improved uncertainty by a factor 3 to 10. 
The most striking contributions come from the large $|t|$, middle $(x_B, Q^2)$ and the low $|t|$, low $Q^2$ regions. 
On the contrary, the ratio of uncertainties is far less striking in the intermediate $|t|$ region since the CFFs are well constrained there, and for large $(x_B, Q^2)$ because of the expected low statistics.

\subsection{Bayesian reweighting procedure}

To assess the impact of the expected new dataset on the knowledge of CFFs without going through a lengthy phase of re-training the ANNs and fitting all the existing and expected measurements, it is possible to use the Bayesian reweighting method. 
In practice, a reweighting coefficient is computed for each ANN CFF replica to assess how compatible this replica is with the newly added dataset. 
The reweighting coefficient $\omega_k$ of the $k$-th replica is obtained by the formula \cite{Ball:2011gg}
\begin{equation}
    \omega_k = \frac{1}{Z} \chi_k^{n-1} \exp(-\chi_k^2 / 2)\,, 
    \label{eq:weights}
\end{equation}
where $n$ is the size of the new dataset represented by the row vector $y$, and $\chi_k$ is the goodness of fit indicator between the $k$-th replica and $y$, given by
\begin{equation}
    \chi_k^2 = (y-y_k) \Sigma^{-1} (y-y_k)^T\,,
\end{equation}
where $y_k$ is the row vector of values of the $k$-th replica at the newly added kinematics, $\Sigma$ the covariance matrix of $y$ and $^T$ denotes matrix transposition.
The multiplicative factor $Z$ introduced in Eq.~\eqref{eq:weights} is obtained by requiring that the sum of all weights $\omega_k$ equals 1. 
To obtain an optimal reweighting, the set of ANN replicas should be able to resolve the $n$-dimensional Gaussian distribution associated with the new dataset $y$ and its covariance matrix $\Sigma$. 
Therefore, reweighting replicas on a vast new dataset spread over a large kinematic region, including regions where the replicas were already poorly constrained, is a difficult task. 
One way of monitoring the statistical relevance of the reweighting procedure relies on the Shannon entropy of the set of weights to evaluate the \textit{effective} number of replicas
\begin{equation}
    N_{\mathrm{eff}} = \exp\left(-\sum_{k=1}^{N_{\textrm{rep}}} \omega_k \log(\omega_k)\right)\,.
\end{equation}
$N_{\mathrm{eff}}$ can be interpreted as the number of replicas effectively carrying information after the introduction of the new dataset.
It should remain reasonably high to ensure a trustworthy statistical representation.

\section{Results and discussion}

\begin{figure}
\includegraphics[scale=0.65]{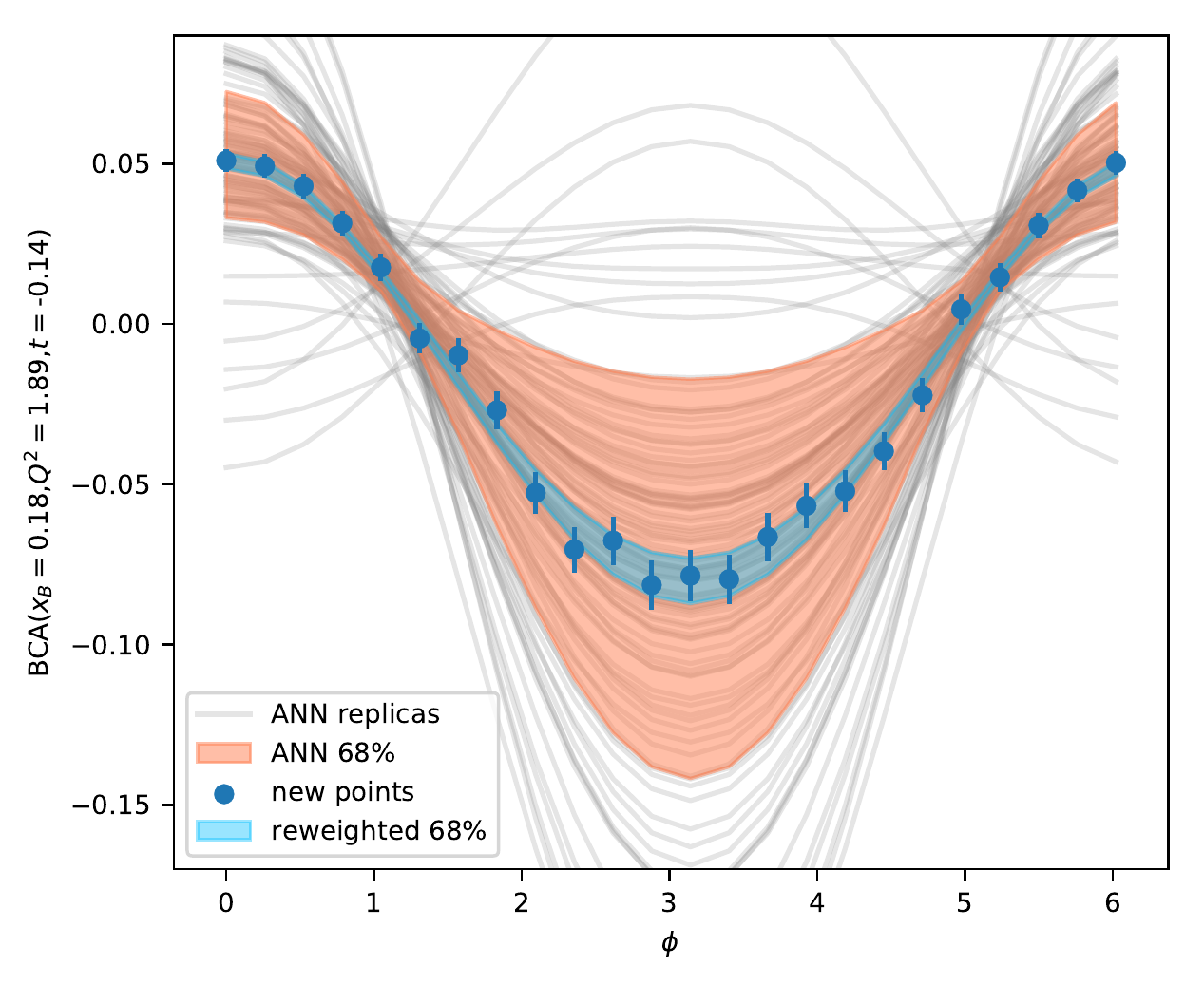}
\caption{\label{fig:localrew} BCA at $(x_B = 0.18, Q^2 = 1.89 \textrm{ GeV}^2, t = -0.14 \textrm{ GeV}^2)$. The light grey lines are the results obtained with the $N_{\textrm{rep}} = 101$ ANN CFF replicas, the red band is the associated 68\% confidence region. The blue dots are the expected experimental points and their uncertainty. The blue band is the reweighted uncertainty. In this case, the number of effective replicas after reweighing is $N_{\mathrm{eff}} = 8$.}
\end{figure}

\begin{figure*}
\includegraphics[scale=0.6]{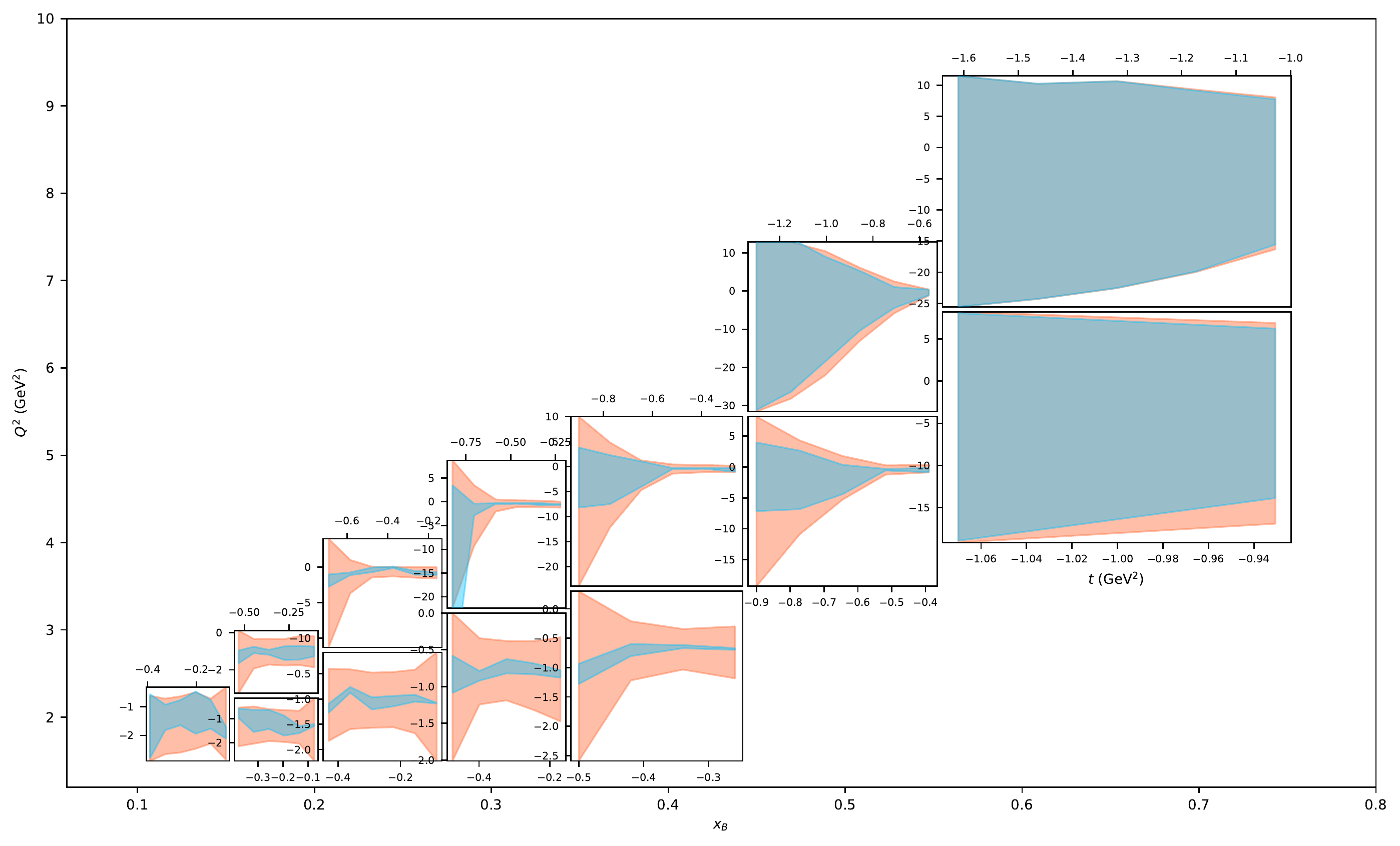}
\caption{\label{fig:localrewH} 68\% confidence regions for $\mathrm{Re}\,\mathcal{H}$. For each $(x_B, Q^2)$ bin, the 68\% confidence regions are shown as a function of $t$ before (red band) and after (blue band) Bayesian reweighting.}
\end{figure*}

To get a flavor of the statistical significance of the expected BCA measurements, we first perform a Bayesian reweighting of the $\phi$-dependence of the BCA distribution computed with the ANN CFFs on a fixed $(x_B, Q^2, t)$ kinematic configuration.
This amounts to neglecting all correlations between the values of a replica at different kinematics $(x_B, Q^2, t)$, but allows us to locally attribute a weight to each replica and compute the consequence on the CFF uncertainty.
\reffig{fig:localrew} shows the impact of such a reweighting of the $\phi$-dependence on the BCA in the bin $(x_B = 0.18, Q^2 = 1.89 \textrm{ GeV}^2, t = -0.14 \textrm{ GeV}^2)$. 
A large reduction of uncertainty is observed. 
The number of replicas still relevant after reweighting has been dramatically reduced from $N_{\textrm{rep}} = 101$ to $N_{\textrm{eff}} = 8$.  
For a few of the $(x_B, Q^2, t)$ bins, the effective number of replicas after reweighting drops as low as about 2. 
It means that the newly added 24 experimental points in $\phi$ are so precise compared to the current uncertainty inherited from ANN CFFs that the weights of all replicas are essentially zero except for a couple of least bad fits. 
In such cases, the result will be sensitive to the exact position of the central values of simulated future experimental points drawn according to Eq.~\eqref{eq:drawcentral} and might not give a fair account of the resulting uncertainty because of the small effective number of replicas used to compute the standard deviation. To prevent exacerbated sensitivity to a specific position of the central values, we average the 68\% uncertainty band on 300 successive smearing of the expected central values. 
This allows us to reinflate a bit the uncertainty band in case of specifically small effective number of replicas, and conversely to reduce abnormally large uncertainty when the effective number of replicas is very large.
The effect for every $(x_B, Q^2, t)$ bins on the real part of the CFF $\mathcal{H}$ is summarized in \reffig{fig:localrewH}. 
The uncertainty is considerably reduced, especially in the region $x_B \in [0.15, 0.45]$ and $Q^2 \in [2, 5]$ GeV$^2$. 

This result already demonstrates the relevance of operating a positron beam at JLab. 
The consensus emerging from the first decade of CFF fits \cite{Kumericki:2016ehc} singles out $\mathrm{Im}\,\mathcal{H}$ as the best determined quantity among all leading twist chiral even CFFs, while $\mathrm{Re}\,\mathcal{H}$ is still poorly known, with even some uncertainty about its sign \cite{Cuic:2020iwt}. 
The definition of the new observables allowed by a positron beam would notably improve our knowledge of $\mathrm{Re}\,\mathcal{H}$. 
An accurate determination of $\mathrm{Re}\,\mathcal{H}$ would have far-reaching consequences. 
First, it could help reduce the current uncertainties in the experimental determination of pressure forces from the DVCS subtraction constant \cite{Dutrieux:2021nlz}.
This subtraction constant can be computed by implementing a dispersion relation with experimental determinations of $\mathrm{Re}\,\mathcal{H}$ and $\mathrm{Im}\,\mathcal{H}$.
Current unbiased determinations of CFFs yield an estimation of this quantity compatible with zero at the 68\% confidence level \cite{Kumericki:2019ddg,Moutarde:2019tqa}. 
This basically prevents any model-independent experimental determination of pressure forces inside the proton.
Although a global fit still seems necessary to actually quantify the gain in the determination of the subtraction constant, a positron beam promises a significant input for such studies.
Second, timelike Compton scattering (TCS), \ie DVCS in the timelike region, is a very promising channel to probe GPDs \cite{Berger:2001xd,Pire:2011st,Moutarde:2013qs,Anikin:2017fwu}. 
It permits a stringent test of leading twist dominance while still working at the level of CFFs.
Indeed CFFs in the spacelike and timelike regions are simply complex-conjugated at leading order and more generally obey simple similar relations that can be systematically checked \cite{Muller:2012yq,Grocholski:2019pqj}.
In particular TCS exhibits a marked dependence on the real part of the (spacelike) CFF $\mathcal{H}$.
The conjunction of BCAs and of various TCS observables would thus allow a highly non-trivial test of the universality of GPDs while still benefiting of a clean electromagnetic probe.

\begin{figure}
\includegraphics[scale=0.7]{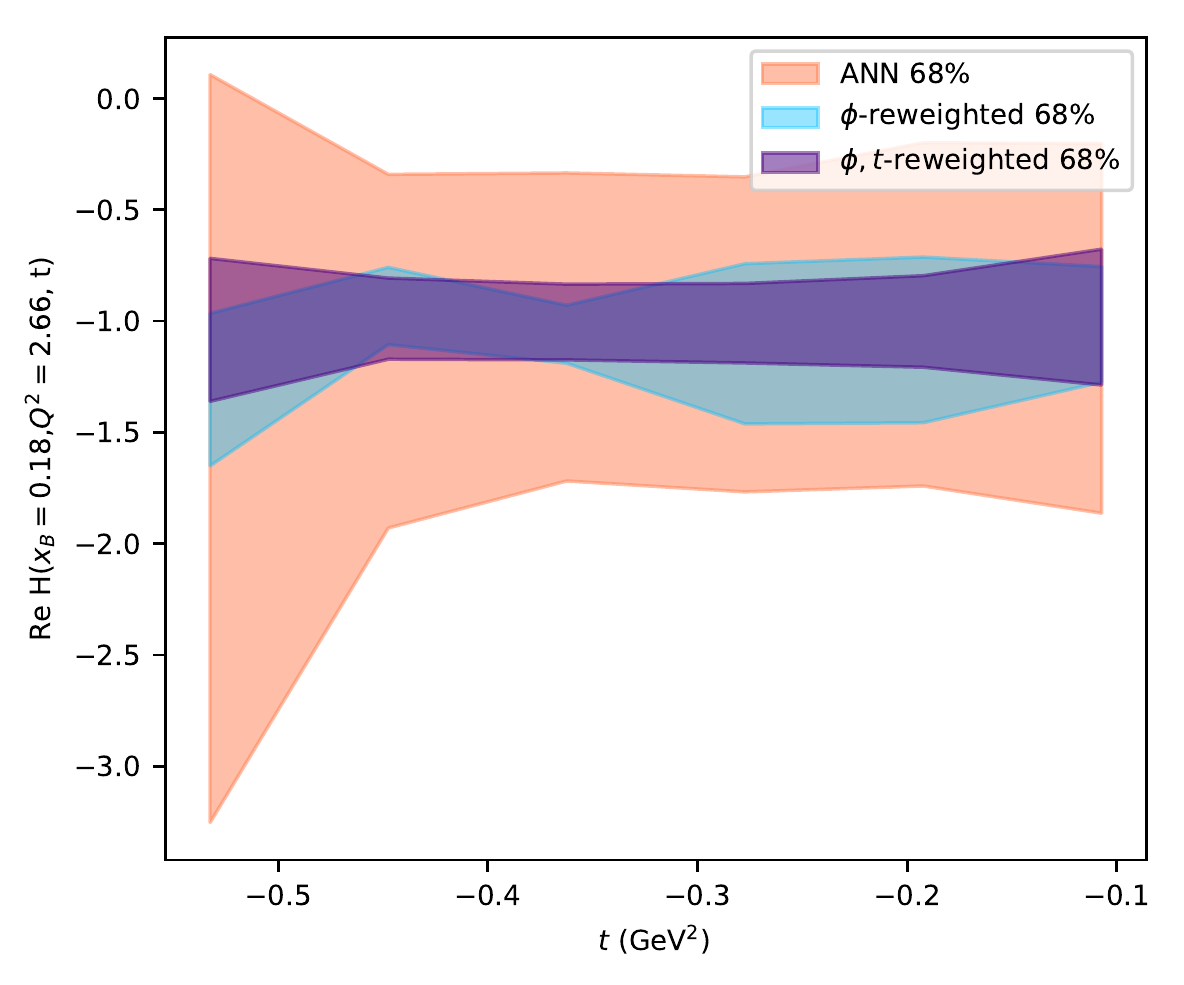}
\caption{\label{fig:globalrewH} 68\% confidence regions for $\mathrm{Re}\,\mathcal{H}$ at $(x_B = 0.18, Q^2 = 2.66\,\textrm{GeV}^2)$. The red band denotes the initial 68\% confidence level coming from ANNs, the blue band results from reweighting the $\phi$-dependence of the BCA at each $(x_B, Q^2, t)$ bin separately, and the purple band results from reweighting the ANN replicas over the 6 $t$ bins while taking the blue band as an experiment-like input on CFFs. The effective number of replicas obtained in the purple band at this $(x_B, Q^2)$ bin is $N_{\textrm{eff}} = 11$.}
\end{figure}

Reweighting the $\phi$-dependence of the BCA results in a local reduction of uncertainty at each $(x_B, Q^2, t)$ bin considered independently. 
This neglects however the fact that ANN replicas of CFFs give information on the distribution of replicas in the $(x_B, Q^2, t)$ space, so that we do not have to consider each bin independently. 
The most straightforward way to assess the influence of the new dataset would be to reweight the replicas globally against the new 1656 experimental points and obtain one weight for each of the 101 replicas. 
This attempt is destined to fail since the effective number of replicas that we would obtain would be 1: the reweighting procedure would merely select the sole replica in least bad agreement with the newly added dataset. 
We can still take advantage of the knowledge of CFF replicas in the $(x_B, Q^2, t)$ space 
by considering that the local reweighting described above gives us experimental-like input on CFFs directly. 
This means that we will now reweight the replica bundle as if the 68\% confidence interval determined previously at each $(x_B, Q^2, t)$ kinematic point were an experimental input on CFFs themselves. 
We demonstrate the result of such a reweighting performed on the 6 $t$ bins for a given $(x_B, Q^2)$ bin in \reffig{fig:globalrewH}.
As expected, exploiting the information of the structure of the replica bundle when varying $(x_B, Q^2, t)$ tends to further decrease the size of the 68\% confidence region. 
However the non-trivial dependence of a CFF replica on kinematic variables can generate less intuitive effects since the first stage of reweighting may have put a local emphasis on different replicas at each $(x_B, Q^2, t)$ kinematic point. 
The proposed secondary reweighting procedure highlights that bin-by-bin reweighting omits some information contained in the replica bundle, but our attempt still remains limited in scope by the achievable statistical significance.

\section{Conclusion}

The difficulty of retrieving GPDs from CFFs --the so-called \emph{deconvolution problem}-- has recently been stressed in Ref.~\cite{Bertone:2021yyz}. 
Waiting for further criterions to bridge the gap between GPD and CFF extractions, it is extremely important to assess which future experimental data (observable, kinematic domain and accuracy) would maximize the improvement of our knowledge of CFFs. 

In this study we used a global fit to existing DVCS measurements where CFFs are described by neural networks. 
These CFFs do not suffer from the systematic uncertainties brought by phenomenological assumptions.
They are the ideal input to statistical studies investigating the impact of new measurements.
To this aim, we conducted a Bayesian reweighting analysis to evaluate the potential of operating a positron beam at JLab and collecting data with CLAS12 for the determination of the real part of the CFF $\mathcal{H}$.
We observe a marked improvement; in some kinematic regions only few data points can reduce the uncertainty on $\mathrm{Re}\,\mathcal{H}$ by a factor about 3.
Data appear to be so constraining that they bring this Bayesian reweighing procedure to its limit of applicability as shown by the small effective number of replicas.

Our present Bayesian reweighting analysis is still uncommon in the GPD domain. 
We demonstrated its applicability in the case of the measurement of beam charge asymmetries in JLab Hall B, but it can of course be straightforwardly extended to other observables, experimental halls in JLab or more generally other facilities. 
Assessing which type of observable should be measured on which channel at which facility has been for a long time a key question of the GPD experimental program. Often the general picture emerges more from a collection of various measurements than from a specific one.
It is hoped that the tools advocated here will provide simpler, faster and more reliable answers to this question.

\begin{acknowledgements}
The authors thank F.~Bossù, M.~Defurne, C.~Mezrag and E.~Voutier for fruitful discussions. This project was supported by the European Union's Horizon 2020 research and innovation programme under grant agreement No 824093. This work was supported by the Grant No. 2019/35/D/ST2/00272 of the National Science Center, Poland. The project is co-financed by the Polish National Agency for Academic Exchange and by the COPIN-IN2P3 Agreement. 
\end{acknowledgements}

\bibliographystyle{spphys}
\bibliography{positrons}

\begin{thebibliography}{10}
\providecommand{\url}[1]{{#1}}
\providecommand{\urlprefix}{URL }
\expandafter\ifx\csname urlstyle\endcsname\relax
  \providecommand{\doi}[1]{DOI \discretionary{}{}{}#1}\else
  \providecommand{\doi}{DOI \discretionary{}{}{}\begingroup
  \urlstyle{rm}\Url}\fi

\bibitem{Mueller:1998fv}
D.~M\"uller, D.~Robaschik, B.~Geyer, F.M. Dittes, J.~Ho\v{r}ej\v{s}i, Fortsch.
  Phys. \textbf{42}, 101 (1994).
\newblock \doi{10.1002/prop.2190420202}

\bibitem{Ji:1996ek}
X.D. Ji, Phys. Rev. Lett. \textbf{78}, 610 (1997).
\newblock \doi{10.1103/PhysRevLett.78.610}

\bibitem{Ji:1996nm}
X.D. Ji, Phys. Rev. D \textbf{55}, 7114 (1997).
\newblock \doi{10.1103/PhysRevD.55.7114}

\bibitem{Radyushkin:1996nd}
A.V. Radyushkin, Phys. Lett. B \textbf{380}, 417 (1996).
\newblock \doi{10.1016/0370-2693(96)00528-X}

\bibitem{Radyushkin:1997ki}
A.V. Radyushkin, Phys. Rev. D \textbf{56}, 5524 (1997).
\newblock \doi{10.1103/PhysRevD.56.5524}

\bibitem{Airapetian:2001yk}
A.~Airapetian, et~al., Phys. Rev. Lett. \textbf{87}, 182001 (2001).
\newblock \doi{10.1103/PhysRevLett.87.182001}

\bibitem{Stepanyan:2001sm}
S.~Stepanyan, et~al., Phys. Rev. Lett. \textbf{87}, 182002 (2001).
\newblock \doi{10.1103/PhysRevLett.87.182002}

\bibitem{Ji:1998pc}
X.D. Ji, J. Phys. G \textbf{24}, 1181 (1998).
\newblock \doi{10.1088/0954-3899/24/7/002}

\bibitem{Goeke:2001tz}
K.~Goeke, M.V. Polyakov, M.~Vanderhaeghen, Prog. Part. Nucl. Phys. \textbf{47},
  401 (2001).
\newblock \doi{10.1016/S0146-6410(01)00158-2}

\bibitem{Diehl:2003ny}
M.~Diehl, Phys. Rept. \textbf{388}, 41 (2003).
\newblock \doi{10.1016/j.physrep.2003.08.002, 10.3204/DESY-THESIS-2003-018}

\bibitem{Belitsky:2005qn}
A.V. Belitsky, A.V. Radyushkin, Phys. Rept. \textbf{418}, 1 (2005).
\newblock \doi{10.1016/j.physrep.2005.06.002}

\bibitem{Boffi:2007yc}
S.~Boffi, B.~Pasquini, Riv. Nuovo Cim. \textbf{30}, 387 (2007).
\newblock \doi{10.1393/ncr/i2007-10025-7}

\bibitem{Guidal:2013rya}
M.~Guidal, H.~Moutarde, M.~Vanderhaeghen, Rept. Prog. Phys. \textbf{76}, 066202
  (2013).
\newblock \doi{10.1088/0034-4885/76/6/066202}

\bibitem{Mueller:2014hsa}
D.~Mueller, Few Body Syst. \textbf{55}, 317 (2014).
\newblock \doi{10.1007/s00601-014-0894-3}

\bibitem{dHose:2016mda}
N.~d'Hose, S.~Niccolai, A.~Rostomyan, Eur. Phys. J. A \textbf{52}(6), 151
  (2016).
\newblock \doi{10.1140/epja/i2016-16151-9}

\bibitem{Accardi:2012qut}
A.~Accardi, et~al., Eur. Phys. J. \textbf{A52}(9), 268 (2016).
\newblock \doi{10.1140/epja/i2016-16268-9}

\bibitem{AbdulKhalek:2021gbh}
R.~Abdul~Khalek, et~al.
\newblock {Science Requirements and Detector Concepts for the Electron-Ion
  Collider: EIC Yellow Report} (2021).
\newblock {arXiv:2103.05419 [physics.ins-det]}

\bibitem{Anderle:2021wcy}
D.P. Anderle, et~al.
\newblock {Electron-Ion Collider in China} (2021).
\newblock {arXiv:2102.09222 [nucl-ex]}

\bibitem{AbelleiraFernandez:2012cc}
J.L. Abelleira~Fernandez, et~al., J. Phys. \textbf{G39}, 075001 (2012).
\newblock \doi{10.1088/0954-3899/39/7/075001}

\bibitem{Kumericki:2016ehc}
K.~Kumericki, S.~Liuti, H.~Moutarde, Eur. Phys. J. A \textbf{52}(6), 157
  (2016).
\newblock \doi{10.1140/epja/i2016-16157-3}

\bibitem{Defurne:2017paw}
M.~Defurne, et~al., Nature Commun. \textbf{8}(1), 1408 (2017).
\newblock \doi{10.1038/s41467-017-01819-3}

\bibitem{Benali:2020vma}
M.~Benali, et~al., Nature Phys. \textbf{16}(2), 191 (2020).
\newblock \doi{10.1038/s41567-019-0774-3}

\bibitem{Kriesten:2020apm}
B.~Kriesten, S.~Liuti.
\newblock {Novel Rosenbluth Extraction Framework for Compton Form Factors from
  Deeply Virtual Exclusive Experiments} (2020).
\newblock {arXiv:2011.04484 [hep-ph]}

\bibitem{Dupre:2016mai}
R.~Dupre, M.~Guidal, M.~Vanderhaeghen, Phys. Rev. D \textbf{95}(1), 011501
  (2017).
\newblock \doi{10.1103/PhysRevD.95.011501}

\bibitem{Dupre:2017hfs}
R.~Dupr\'e, M.~Guidal, S.~Niccolai, M.~Vanderhaeghen, Eur. Phys. J. A
  \textbf{53}(8), 171 (2017).
\newblock \doi{10.1140/epja/i2017-12356-8}

\bibitem{Burkert:2018bqq}
V.D. Burkert, L.~Elouadrhiri, F.X. Girod, Nature \textbf{557}(7705), 396
  (2018).
\newblock \doi{10.1038/s41586-018-0060-z}

\bibitem{Burkert:2021ith}
V.D. Burkert, L.~Elouadrhiri, F.X. Girod.
\newblock {Determination of shear forces inside the proton} (2021).
\newblock {arXiv:2104.02031 [nucl-ex]}

\bibitem{Moutarde:2018kwr}
H.~Moutarde, P.~Sznajder, J.~Wagner, Eur. Phys. J. C \textbf{78}(11), 890
  (2018).
\newblock \doi{10.1140/epjc/s10052-018-6359-y}

\bibitem{Kumericki:2019ddg}
K.~Kumeri\v{c}ki, Nature \textbf{570}(7759), E1 (2019).
\newblock \doi{10.1038/s41586-019-1211-6}

\bibitem{Cuic:2020iwt}
M.~\v{C}ui\'c, K.~Kumeri\v{c}ki, A.~Sch\"afer, Phys. Rev. Lett.
  \textbf{125}(23), 232005 (2020).
\newblock \doi{10.1103/PhysRevLett.125.232005}

\bibitem{Moutarde:2019tqa}
H.~Moutarde, P.~Sznajder, J.~Wagner, Eur. Phys. J. C \textbf{79}(7), 614
  (2019).
\newblock \doi{10.1140/epjc/s10052-019-7117-5}

\bibitem{Afanasev:2019xmr}
A.~Afanasev, et~al.
\newblock {Physics with Positron Beams at Jefferson Lab 12 GeV} (2019).
\newblock {arXiv:1906.09419 [nucl-ex]}

\bibitem{Accardi:2020swt}
A.~Accardi, et~al.
\newblock {e$^+$@JLab White Paper: An Experimental Program with Positron Beams
  at Jefferson Lab} (2020).
\newblock {arXiv:2007.15081 [nucl-ex]}

\bibitem{Burkert:2020akg}
V.D. Burkert, et~al., Nucl. Instrum. Meth. A \textbf{959}, 163419 (2020).
\newblock \doi{10.1016/j.nima.2020.163419}

\bibitem{Ball:2010gb}
R.D. Ball, V.~Bertone, F.~Cerutti, L.~Del~Debbio, S.~Forte, A.~Guffanti, J.I.
  Latorre, J.~Rojo, M.~Ubiali, Nucl. Phys. B \textbf{849}, 112 (2011).
\newblock \doi{10.1016/j.nuclphysb.2011.03.017}.
\newblock [Erratum: Nucl.Phys.B 854, 926--927 (2012), Erratum: Nucl.Phys.B 855,
  927--928 (2012)]

\bibitem{Ball:2011gg}
R.D. Ball, V.~Bertone, F.~Cerutti, L.~Del~Debbio, S.~Forte, A.~Guffanti, N.P.
  Hartland, J.I. Latorre, J.~Rojo, M.~Ubiali, Nucl. Phys. B \textbf{855}, 608
  (2012).
\newblock \doi{10.1016/j.nuclphysb.2011.10.018}

\bibitem{Lin:2017snn}
H.W. Lin, et~al., Prog. Part. Nucl. Phys. \textbf{100}, 107 (2018).
\newblock \doi{10.1016/j.ppnp.2018.01.007}

\bibitem{Dutrieux:2021nlz}
H.~Dutrieux, C.~Lorc\'e, H.~Moutarde, P.~Sznajder, A.~Trawi\'nski, J.~Wagner,
  Eur. Phys. J. C \textbf{81}(4), 300 (2021).
\newblock \doi{10.1140/epjc/s10052-021-09069-w}

\bibitem{Kroll:2012sm}
P.~Kroll, H.~Moutarde, F.~Sabatie, Eur. Phys. J. C \textbf{73}(1), 2278 (2013).
\newblock \doi{10.1140/epjc/s10052-013-2278-0}

\bibitem{Belitsky:2012ch}
A.V. Belitsky, D.~M\"uller, Y.~Ji, Nucl. Phys. B \textbf{878}, 214 (2014).
\newblock \doi{10.1016/j.nuclphysb.2013.11.014}

\bibitem{burkert:pac48}
V.D. Burkert, et~al.
\newblock {Beam Charge Asymmetries for Deeply Virtual Compton Scattering on the
  Proton at CLAS12} (2020).
\newblock {Proposal to PAC48}

\bibitem{Berger:2001xd}
E.R. Berger, M.~Diehl, B.~Pire, Eur. Phys. J. \textbf{C23}, 675 (2002).
\newblock \doi{10.1007/s100520200917}

\bibitem{Pire:2011st}
B.~Pire, L.~Szymanowski, J.~Wagner, Phys. Rev. \textbf{D83}, 034009 (2011).
\newblock \doi{10.1103/PhysRevD.83.034009}

\bibitem{Moutarde:2013qs}
H.~Moutarde, B.~Pire, F.~Sabatie, L.~Szymanowski, J.~Wagner, Phys. Rev.
  \textbf{D87}(5), 054029 (2013).
\newblock \doi{10.1103/PhysRevD.87.054029}

\bibitem{Anikin:2017fwu}
I.V. Anikin, et~al., Acta Phys. Polon. \textbf{B49}, 741 (2018).
\newblock \doi{10.5506/APhysPolB.49.741}

\bibitem{Muller:2012yq}
D.~Mueller, B.~Pire, L.~Szymanowski, J.~Wagner, Phys. Rev. \textbf{D86}, 031502
  (2012).
\newblock \doi{10.1103/PhysRevD.86.031502}

\bibitem{Grocholski:2019pqj}
O.~Grocholski, H.~Moutarde, B.~Pire, P.~Sznajder, J.~Wagner, Eur. Phys. J. C
  \textbf{80}(2), 171 (2020).
\newblock \doi{10.1140/epjc/s10052-020-7700-9}

\bibitem{Bertone:2021yyz}
V.~Bertone, H.~Dutrieux, C.~Mezrag, H.~Moutarde, P.~Sznajder.
\newblock {The deconvolution problem of deeply virtual Compton scattering}
  (2021).
\newblock {arXiv:2104.03836 [hep-ph]}

\end{thebibliography}

\end{document}